\documentclass{article}

\usepackage{enumitem}
\usepackage{graphicx}
\usepackage{dcolumn}
\usepackage{booktabs}
\usepackage{imakeidx}
\usepackage{graphics}
\usepackage{array}
\usepackage{bm}
\usepackage{color}
\usepackage{amsmath}
\usepackage{longtable}
\input{epsf.tex}
\def\be{\begin{equation}}
\def\ee{\end{equation}}
\def\ba{\begin{array}}
\def\ea{\end{array}}
\def\bea{\begin{eqnarray}}
\def\eea{\end{eqnarray}}

\begin{document}
\renewcommand\arraystretch{1.5}

\newcommand{\BA}[1]{\textcolor{red}{BA: #1}}
\newcommand{\av}[1]{\textcolor{blue}{AV: #1}}
\setlength{\abovecaptionskip}{0.1cm}
\setlength{\belowcaptionskip}{0.5cm}
\pagestyle{empty}
\newpage
\begin{center} {\large\bf Estimation of the slope of nuclear symmetry energy via charge radii of mirror nuclei}\\
\vspace*{0.4cm} {Sakshi Gautam$^{1,2}$}, {Anagh Venneti$^{2}$}, {Sarmistha Banik $^{2}$} \\ and {B. K. Agrawal$^{3}$}\\
{\it $^{1}$ Department of Physics, Panjab University, Chandigarh -160014, India.\\}
{\it $^{2}$ Department of Physics, Birla Institute of Technology and Science Pilani, Hyderabad Campus, Hyderabad-
500078, India. \\}
{\it $^{3}$ Saha Institute of Nuclear Physics, 1/AF Bidhan Nagar, Kolkata-700064, India. \\}
Email:~sakshigautam@pu.ac.in \\

\end{center}

\begin{abstract}
Charge radii of mirror nuclei are calculated by implementing pairing effects with the Hartree-Fock Bogoliubov approximation. Correlations between the difference of charge radii ($\Delta R_{ch}$) and slope of nuclear symmetry energy (L) are examined for different mirror nuclei pairs of varying masses using 40 different Skyrme energy density functionals. $\Delta R_{ch}-L $ correlations are found to be robust for the binding constraints imposed on density functionals. We observe that $\Delta R_{ch}$ and $L$ show better correlations in relatively heavier pairs than those obtained in the lighter pairs. Our calculations impose a constraint on the slope of nuclear symmetry energy as -20 MeV $\leq L \leq$ 55 MeV with 68\% confidence band using available measurements on charge radii. This is a moderately soft symmetry energy, in contrast to stiff and soft symmetry energy indicated by PREX-II and CREX measurements of neutron skin thickness in $^{208}Pb$ and $^{48}Ca$, respectively. Our result is also in agreement with celestial constraints obtained from observational data for neutron stars.

\end{abstract}

\section{Introduction}
Energy per particle for asymmetric nuclear matter can be expressed approximately as \cite{baldo}; 

\begin{equation}
\label{eq1}
    E (\rho,\delta) = E(\rho, 0) + E_{sym} (\rho)\delta^2,
\end{equation}
where $E (\rho,0)$ is the energy of symmetric nuclear matter, $\delta = \frac{\rho_n-\rho_p}{\rho_n+\rho_p}$ is the isospin asymmetry parameter and $\rho_{n/p}$ is the density of neutrons/protons, respectively. $E_{sym}$ is the nuclear symmetry energy, which represents the energy difference between pure neutron matter ($\delta = 1$) and symmetric nuclear matter ($\delta = 0$). It measures the change in the nuclear binding of the system as the neutron-to-proton ratio is varied at a fixed baryonic number. The nuclear symmetry energy around saturation density $\rho_0$ is expanded up to second order  as;

\begin{eqnarray}
\label{eq2}
    E_{sym} (\rho) &= &J + L \left(\frac{\rho - \rho_0}{3 \rho_0}\right) + \frac{1}{2} K_{sym}\left(\frac{\rho-\rho_0}{3 \rho_0}\right)^2,\\
\text{where}&&\nonumber \\
    J &= &E_{sym}(\rho_0),\\
    L & =& 3 \rho_0 \left(\frac{\partial E_{sym}}{\partial \rho}\right)_{\rho=\rho_0},\\
    K_{sym} &= & 9 \rho_0^2 \left(\frac{\partial ^2 E_{sym}}{\partial \rho^2}\right)_{\rho=\rho_0}.
\end{eqnarray}
where $J$ is the nuclear symmetry energy at saturation density, L is the slope of nuclear symmetry energy and $K_{sym}$ is the curvature of nuclear symmetry energy. The slope of nuclear symmetry energy (L) is estimated from the pressure of pure neutron matter at saturation density and is related as;
\begin{equation}
\label{eq3}
     P_N (\rho_0) \approx  \frac{\rho_0 L}{3}.
\end{equation}
The value of $J$ is found to be close to 32 MeV from various studies on nuclear structure \cite{baldo}. However, it is the slope of nuclear symmetry energy (L), which is crucial to understand the extrapolation of symmetry energy to lower and higher densities \cite{brow1,bomb}. A correct estimation of L is important to study the properties of proton- or neutron-rich heavy nuclei as well as to understand the physics of neutron stars.

Strong interactions play a significant role for both neutron star matter and the finite nuclei. Neutron degeneracy pressure acts against inward gravitational pull in the former, unlike in the finite nucleus, wherein it acts against surface tension. In finite nuclei, a higher value of L, due to an increase in neutron pressure, will push the neutron surface (see eqn. \ref{eq3}), thus leading to an enhanced neutron distribution in neutron-rich nuclei. Therefore, the slope of symmetry energy determines neutron-skin thickness, \emph{viz., the difference in the root mean square (r.m.s.) radii of neutrons and protons distributions},  in finite nuclei as well as the radius of neutron stars. Note that curvature of symmetry energy, $K_{sym}$, also plays a significant role in astrophysical processes, where high densities are involved and its value is poorly constrained at present \cite{bali2,holt}. Thus, multi-messenger observations from  the merging of neutron stars provide excellent probes of high-density nuclear matter to complement the knowledge gained from terrestrial measurements

\par
Since, $L$ is not a directly measurable quantity, various probes of L have been proposed in recent times from the studies involving neutron skins in finite nuclei \cite{brow1,reinh,thei,suzu,chen}, collective flows, and fragmentation observables in heavy-ion collisions \cite{wang1,lync,tsan,russ}. In addition, astrophysical observations of Gravitational waves relating to tidal deformability and radius of neutron stars \cite{oert,baio,rait,bali} have been employed to constrain the values of $L$.  The values of $L$ estimated from different observables are usually at variance, thus, the density dependence of symmetry energy remains inconclusive.  Neutron skin thickness for $Pb^{208}$ and $Ca^{48}$ nuclei have been measured by  PREX-I/II \cite{adhip,abra, horo} and CREX \cite{adhic} experiments, respectively, to obtain constraints on the values of $L$. For example, improved neutron skin thickness of $Pb^{208}$ by  PREX-II  hints towards a stiffer symmetry energy.
 The strong correlation between neutron skin thickness in $Pb^{208}$ and $L$  yields $L = 106 \pm$37 within the relativistic energy density functional \cite{reed}. On the other hand, neutron skin thickness for $Ca^{48}$ nucleus from CREX results in much softer symmetry energy with L = 30.61 $\pm$ 6.74 MeV \cite{kuma} with relativistic energy density functionals. This disagreement in the simultaneous reproduction of both measurements reflects an incomplete knowledge of the equation of the state of neutron-rich matter. Similar conclusions of marginal overlap of L values are reported by Tagami et al. \cite{taga} with around 200 equations of state. They proposed L values for CREX and PREX-II results to be between 0-51 and 76-165 MeV, respectively. Likewise, a tension between CREX and PREX results is recently reported in Ref. \cite{zhang}, where Bayesian inference of nuclear symmetry energy and neutron skins in $Pb^{208}$ and $Ca^{48}$ is performed. The values of symmetry energy are inferred separately from CREX and PREX-II, compatible with each other at 90$\%$ confidence interval and inconsistent with each other at 68.3$
\%$  confidence interval. Similar inadequacy of relativistic energy density functionals in the simultaneous understanding of CREX and PREX-II results is reported in Ref. \cite{miya} as well. Therefore, there is a large stimulus in the theoretical nuclear physics community to further investigate the slope of nuclear symmetry energy.

 \par
The advancements in laser beam technology result in the precise measurements of nuclear charge radii \cite{yang1} and thus open new opportunities in estimating neutron-skin thicknesses, under isospin symmetry of nucleon-nucleon interaction. At the same time, measuring neutron radii is a bit challenging and involves sensitive electro-weak interactions. An alternative clean electromagnetic probe, \emph{viz}., charge radii in mirror nuclei, is proposed to estimate the slope of nuclear symmetry energy \cite{brow2}. It is based on the assumption that under perfect charge symmetry, the neutron radius in a given nucleus is the same as the proton radius in its mirror counterpart. Therefore, neutron skin thickness ($\Delta R_{np}$) in a nucleus is equivalent to the difference in proton radii ($\Delta R_{ch}$) of mirror nuclei pairs (under no Coulomb interactions) \cite{brow2, yang}; i.e.,
\begin{equation}
    R_n (X^N_Z) - R_p (X^N_Z)  = R_p (Y^Z_N) - R_p (X^N_Z),
\end{equation}
where $X^N_Z$ and $Y^Z_N$ are mirror nuclei pairs. As mentioned above, the foremost attempt was carried out by Brown for the mirror nuclei pairs of $^{34}\textrm{Ar} $-$^{34}$\textrm{S}, $^{52}\textrm{Ni} $-$^{52}$\textrm{Cr} and $^{54}\textrm{Ni} $-$^{54}$\textrm{Fe} using 48 Skyrme energy density functionals \cite{brow2}. It was shown that the difference in charge radii is proportional to $(N-Z)\times L$. Later, this linear correlation between $\Delta R_{ch}$ and L was also confirmed by Yang and Piekarewicz \cite{yang}, using relativistic density functionals. Following these studies, constraints on L are established using recent precise measurements on radii of proton-rich nuclei of $^{36}$\textrm{Ca}/$^{38}$\textrm{Ca} and $^{54}$\textrm{Ni}. The deduced values of L are 5-70 MeV \cite{brow3} and 21-88 MeV \cite{pine}, respectively. Another very recent attempt in this direction was carried out by An \emph {et al.} \cite{an}, where $\Delta R_{ch}$ and L correlations are observed for different pairs using six relativistic and six non-relativistic energy density functionals. Their deduced slope values from comparison with measured radii are reported to be between 22.5 and 51.55 MeV. Recently, correlations between charge radii differences of mirror nuclei and neutron star observables are also investigated \cite{pbano}. Therefore, we see that a lot of studies on charge radii calculations are still in progress for obtaining a stringent constraint on L. However, this claim of establishing a clean probe of L was challenged in a study by Reinhard and Naraewicz \cite{rein2}, wherein it is observed that $\Delta R_{ch}$ is an inferior indicator of L as it is influenced by pairing correlations in nuclei. Lately, the above finding was verified by Huang \emph {et al}. \cite{huang} for different mirror nuclei pairs for around 20 Skyrme functionals. These investigations thus pointed out that even precise measurements of $\Delta R_{ch}$ cannot put any stringent constraints on the slope of the nuclear symmetry energy.

\par
In the present work, we shall re-visit the $\Delta R_{ch}$ $vs$ L correlation in light of pairing effects using 40 Skyrme energy density functionals. We shall also examine the robustness of these correlations by further considering only those density functionals that predict nuclear binding energies to be within $\pm 10$ MeV of the measured nuclear bindings. Further, relatively better correlations of $\Delta R_{ch}$ $vs$ L are seen in heavier nuclei, and possible bounds on L are reported. A brief outline of the theoretical framework is presented in Section II and results are discussed in Section III, with a summary in Section IV.  

\section{Theoretical approach}
\par The present investigation of the correlation between the difference in charge radii of mirror nuclei and the slope of nuclear symmetry energy is carried out for 40 Skyrme density functionals, for which L varies between approximately -30 MeV to 160 MeV. This wide range of L covers most of the values of L employed in previous studies. Various Skyrme forces used in this work are listed in Table \ref{table:1}. The values of saturation density, binding energy per nucleon at saturation, incompressibility coefficient, symmetry energy coefficient, slope of symmetry energy, and curvature of symmetry energy for various functionals are also listed in the Table. 
\par
The nuclei involved in the present study are open-shell nuclei, therefore, pairing effects will be significant and are treated using Hartree-Bogoliubov transformations \cite{ring}.  The Hartree-Fock-Bogoliubov equation for nucleons is:
\begin{equation}
  \begin{pmatrix}
      h & \Delta\\
      -\Delta^* & -h^*
  \end{pmatrix}
  \begin{pmatrix}
      U_i\\
      V_i
  \end{pmatrix}
  = E_i
  \begin{pmatrix}
      U_i\\
      V_i
  \end{pmatrix}
\end{equation}
in which $E_i$ is the quasi-particle energy, $(U_i,V_i)$ is the quasi-particle wave function and $\emph{h}$ is the single-particle Hamiltonian of the nucleon, which comprises contributions from the kinetic term, mean-field term, and Fermi energy. Here $\Delta$ represents the pairing potential:
\begin{equation}
    \Delta_{jk} = \frac{1}{2} \sum_{lm} \langle jk \vert V^{n,p} \vert lm \rangle \lambda_{lm}
\end{equation}
where $\lambda$ is the pairing tensor and $V^{n,p}$ is the pairing force.
The pairing channel is parameterized by a density-dependent delta-pairing force with mixed volume and surface features, as:
 \begin{equation}
  V^{n,p}_{pair}(r) =  V^{n,p}_0(r) \left(1-\frac{1}{2}\frac{\rho(r)}{\rho_0}\right) \delta (r-r^{'}),   
 \end{equation}
 with $V^{n,p}_0(\bf r)$ the pairing strength for neutrons (n) and protons (p) (specific for each density functional), $\rho(\bf r)$ the isoscalar local density. As the pairing force introduced is of zero range, a cut-off in the quasi-particle space is introduced, which we take as 60 MeV in the present calculations.
 \par
The charge radii ($R_{ch}$) of different nuclei are calculated using the relation:
 \begin{equation}
     R_{ch} = \sqrt{R_p^2 + 0.8^2}.
 \end{equation}
 Here, $R_p$ is the point proton radius, and the factor of 0.8 accounts for finite-size correction \cite{lala,buch}. The correlation between the difference of charge radius in mirror nuclei and the slope of symmetry energy is analyzed through least-square analysis and is quantified using the Pearson correlation coefficient. 
 
\begin{center}
\begin{longtable}{|c|c|c|c|c|c|c|c|}
\caption {Values of saturation density ($\rho_0$), binding energy per nucleon at saturation density ($E_0$), incompressibility coefficient (K), symmetry energy coefficient (J), slope of symmetry energy (L)  and curvature of symmetry energy (K$_{sym}$) for various Skyrme energy density functionals. All the quantities are in units of  MeV, except for $\rho_0$ which is in fm$^{-3}$.}.\label{table:1}\\
\hline 
\textrm{Force} & \textrm{$\rho_0$ } & \textrm{$E_0$} & \textrm{K} & \textrm{J} &  \textrm{L} &  \textrm{$K_{sym}$} & \textrm{Ref.} \\
\hline 
 Zs & 0.16 & -15.86 & 233.12 & 26.71 & -29.27 & -401.43 & \cite{zs}\\ 
\hline 
 MSK3 & 0.157 & -15.79 & 233.24 & 28.00 & 7.04 & -283.52 & \cite{msk1}\\
\hline 
 BSK1 & 0.157 & -15.80 & 231.29 & 27.81 & 7.19  & -281.83 & \cite{bsk1}\\ 
\hline 
MSK4 & 0.157 & -15.79 & 231.16 & 28.00 & 7.20 & -284.05 & \cite{msk1} \\
\hline 
MSK8 & 0.157 & -15.79 & 229.30 & 27.93 & 8.26 & -280.01 & \cite{msk7}  \\ 
\hline 
 MSK7 & 0.157 & -15.79 & 231.21 & 27.95 & 9.41 & -274.63 & \cite{msk7}\\
\hline 
MSK6 & 0.157 & -15.79 & 231.16 & 28.00 & 9.63 & -274.33 & \cite{msk1}  \\ 
\hline 
 SIII & 0.145 & -15.85 & 355.35 & 28.16 & 9.91 & -393.73 & \cite{siii} \\
\hline 
 MSK9 & 0.157 & -15.80 & 233.31 & 27.99 & 10.37 & -270.23 & \cite{msk7}  \\ 
\hline 
 SKP & 0.162 & -15.95 & 201.02 & 30.00 & 19.67 & -266.60 & \cite{skp} \\
\hline 
 SKX & 0.155 & -16.06 & 271.11 & 31.09 & 33.17 & -252.12 & \cite{skx} \\ 
\hline 
 MSK2 & 0.157 & -15.83 & 231.63 & 30.00 & 33.35 & -203.44 & \cite{msk1} \\
\hline 
 SKXC & 0.155 & -15.86 & 268.16 & 30.20 & 33.6 & -238.39 & \cite{skx}  \\ 
\hline 
 MSK1 & 0.157 & -15.83 & 233.72 & 30.00 & 33.9 & -200.02 & \cite{msk1} \\
\hline 
 SGII & 0.158 & -15.59 & 214.63 & 26.83 & 37.6 & -145.90 & \cite{sgii} \\ 
\hline 
 HFB9 & 0.159 & -15.92 & 231.42 & 30.00 & 39.89 & -150.03 & \cite{hfb9}  \\
\hline 
 UNE1 & 0.159 & -15.80 & 220.00 & 28.99 & 40.00 & -176.51 & \cite{une1}  \\ 
\hline 
 KDE & 0.164 & -15.97 & 223.69 & 31.97 & 41.4 & -141.83 & \cite{kde} \\
\hline 
UNE0 & 0.160 & -16.05 & 229.99 & 30.54 & 45.08 & -185.74 & \cite{une0}\\ 
\hline 
KDE0 & 0.161 & -16.08 & 228.49 & 32.99 & 45.23 & -144.78 & \cite{kde}  \\
\hline 

 SKM* & 0.160 & -15.78 & 216.66 & 30.03 & 45.77 & -155.94 & \cite{skms} \\ 
\hline 
 SLY4 & 0.159 & -15.97 & 229.90 & 32.00 & 45.96 & -119.73 & \cite{sly4} \\
\hline 
SLY7 & 0.158 & -15.90 & 229.74 & 31.99 & 46.94 & -114.34 & \cite{sly4}  \\ 
\hline 
 SLY6 & 0.159 & -15.92 & 229.84 & 31.96 & 47.45 & -112.71 & \cite{sly4} \\
\hline 
 SKb & 0.155 & -15.99 & 263.14 & 23.88 & 47.5 & -78.46 & \cite{ska}  \\ 
\hline 
SLY5 & 0.161 & -15.98 & 229.91 & 32.01 & 48.15 & -112.76 & \cite{sly4}\\
\hline  
 KDE1 & 0.164 & -16.21 & 227.33 & 34.58 & 54.7 & -127.12 & \cite{kde}  \\ 
\hline 
 SKT1 & 0.161 & -15.98 & 236.11 & 32.01 & 56.2  & -134.83 & \cite{skt1}\\
\hline 
 SKI4 & 0.159 & -15.93 & 235.48 & 35.45 & 60.43 & -40.56 & \cite{ski1} \\ 
\hline 
 SIV & 0.151 & -15.96 & 324.53 & 31.22 & 63.49 & -136.72 & \cite{siii} \\
\hline 
 SKOP & 0.160 & -15.75 & 222.31 & 31.94 & 68.9 & -78.83 & \cite{sko} \\ 
\hline 
 SKa & 0.155 & -15.99 & 263.14 & 32.91 & 74.6  & -78.46 & \cite{ska}\\
\hline 
SKO & 0.160 & -15.83 & 223.32 & 31.97 & 79.14 & -43.17 & \cite{sko} \\
\hline 
 S272 & 0.155 & -16.28 & 271.49 & 37.39 & 91.67 & -67.79 & \cite{s272}  \\ 
\hline 
 SKT4 & 0.159 & -15.94 & 235.48 & 35.45 & 94.1 & -24.46 & \cite{skt1} \\
\hline 
 S255 & 0.157 & -16.33 & 254.92 & 37.39 & 95.05 & -58.34 & \cite{s272} \\ 
\hline 
 SKI3 & 0.157 & -15.96 & 257.97 & 34.83 & 100.49 & 73.04 & \cite{ski1} \\
\hline 
 SKI2 & 0.157 & -15.73 & 240.37 & 33.33 & 104.16 & 70.69 &\cite{ski1}  \\ 
\hline 
 SKI5 & 0.155 & -15.83 & 255.57 & 36.61 & 129.17 & 159.97 & \cite{ski1}  \\
\hline 
SKI1 & 0.161 & -15.98 & 242.53 & 37.51 & 160.96 & 234.67 & \cite{skt1} \\ 
\hline 
\end{longtable}
\end{center}

\section{Results and discussion}
 The present study is carried out for five different pairs of  mirror nuclei: $^{14}$\textrm{O}$ -$$^{14}$\textrm{C},
 $^{18}$\textrm{Ne}$ -$$^{18}$\textrm{O},
 $^{44}$\textrm{Cr}$ -$$^{44}$\textrm{Ca},
 $^{58}$\textrm{Zn}$ -$$^{58}$\textrm{Ni} and $^{60}$\textrm{Ge}$ -$$^{60}$\textrm{Ni}. The mass of these pairs ranges from 14 to 60 units, and this range covers the mass units for which experimental measurements of charge radii are also available. As mentioned previously, the present investigation is done for 40 different Skyrme energy density functionals taking into account the pairing correlations as discussed in Section 2.

\begin{figure}[!t] \centering \vskip 0.0cm
\includegraphics[scale=0.6]{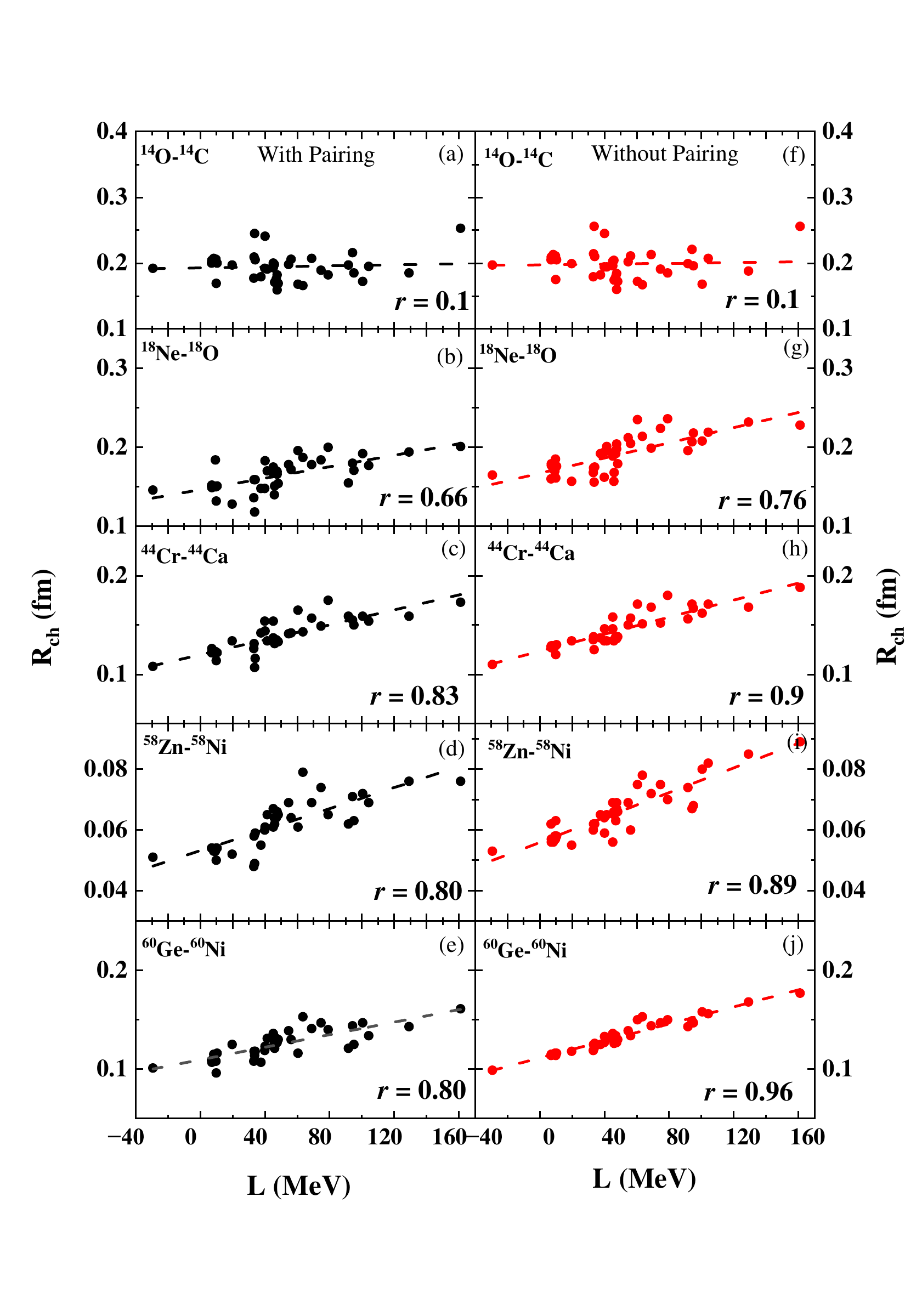}
\caption{$\Delta R_{ch}$ as a function for L for the pairs of
$^{14}$\textrm{O} $-$$^{18}$\textrm{C},[a \& f] $^{18}$\textrm{Ne} $-$$^{18}$\textrm{O} [b \& g], $^{44}$\textrm{Cr} $-$ $^{44}$\textrm{Ca} [c \& h], $^{58}$\textrm{Zn} $-$ $^{58}$\textrm{Ni} [d \& i] and $^{60}$\textrm{Ge} $-$ $^{60}$\textrm{Ni} [e \& j]. Left and right panels display the results with and without pairing effects. The dashed lines represent least-square linear fits and the corresponding Pearson correlation coefficients are mentioned in the panels.} \label{Fig1}
\end{figure}

\par
We have performed the calculations for several nuclei considered with and without the inclusion of the pairing term. 
In Fig. \ref{Fig1}, the difference in the charge radius of mirror nuclei as a function of the slope of symmetry energy are plotted as solid circles obtained with pairing (left panels) and without pairing term (right panels) for five pairs of mirror nuclei. The dashed lines represent the least-square linear fits.  From the figure, we notice that there is almost no correlation for the $^{14}$\textrm{O} $-$ $^{14}$\textrm{C} pair. However, correlations significantly improve for the rest of the pairs of mirror nuclei considered, as reflected in the values of Pearson correlation coefficients which are 0.66, 0.83, 0.8 and 0.8 for $^{18}$\textrm{Ne} $-$$^{18}$\textrm{O}, $^{44}$\textrm{Cr} $-$ $^{44}$\textrm{Ca}, $^{58}$\textrm{Zn} $-$ $^{58}$\textrm{Ni} and $^{60}$\textrm{Ge} $-$ $^{60}$\textrm{Ni}, respectively. 
\par 
Recent studies in Refs. \cite{rein2,huang} have shown that the inclusion of pairing effects weakens the correlation between the difference in charge radii of mirror nuclei and L. To verify this aspect here, we have performed the calculations without taking into account the pairing effects. This is achieved by setting the pairing strength to zero and the corresponding results are shown in the right panels of Fig. 1. Once again a  poor correlation is found in $^{14}$\textrm{O} $-$ $^{14}$\textrm{C} pair. The value of Pearson correlation coefficients now enhance and are 0.76, 0.9, 0.89, and 0.96 for $^{18}$\textrm{Ne} $-$$^{18}$\textrm{O}, $^{44}$\textrm{Cr} $-$ $^{44}$\textrm{Ca}, $^{58}$\textrm{Zn} $-$ $^{58}$\textrm{Ni} and $^{60}$\textrm{Ge} $-$ $^{60}$\textrm{Ni}, respectively.  Our findings thus establish the results reported in Refs. \cite{rein2,huang}. It is worth mentioning that correlations reported in the findings of Ref. \cite{huang} are different than those obtained in the present study. This could be due to different Skyrme energy density functionals used in both studies and different treatments of pairing force (only surface contribution taken into account in Ref. \cite{huang}). Moreover, $\Delta R_{ch}$ and L correlations are also affected by the correlation between the curvature of symmetry energy $K_{sym}$ and L \cite{huang}, which is different in both studies. This aspect will be discussed later in Fig 3.   It is worth mentioning that pairing force also plays a significant role in the static and dynamics aspects of fission phenomena experienced by heavy atomic nuclei as well \cite{zhao,wang}. For example, pairing correlations have a strong impact on the lifetime of spontaneous fission \cite{zhao} as well as on fission yields of $Pu^{240}$ \cite{wang}. 

\begin{figure}[!t] \centering \vskip 0.0cm
\includegraphics[scale=0.6]{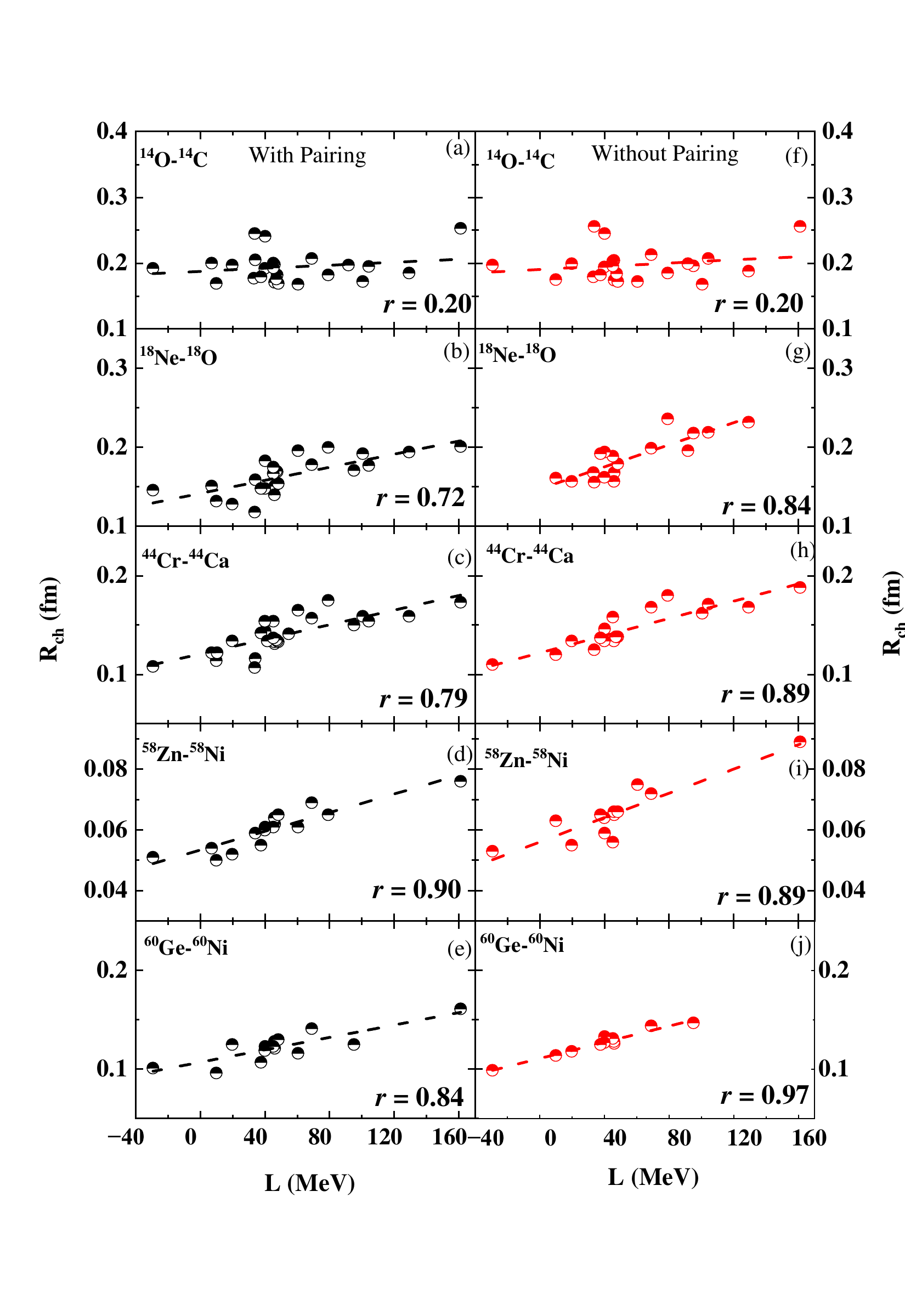}
\caption{Same as Fig. 1, but for density functionals with binding energy constraint of $\pm 10$ MeV (for details, see text). } \label{Fig2}
\end{figure}

\par
Next, we assess the role of pairing effects on the robustness of these correlations. For this, we re-examine $\Delta R_{ch}$ $vs$ L correlation for 
Skyrme forces that reproduce the nuclear bindings of corresponding nuclei to be within $\pm  10$ MeV of experimental values. The data on nuclear bindings is taken from AME2020 atomic mass table \cite{ame}. The imposition of binding energy constraint filters out various Skyrme forces. The results with binding energy constraints are displayed in Fig. \ref{Fig2}. We observe that $\Delta R_{ch}$ $vs$ L correlations do not change significantly with this imposed constraint of binding energy. A poor correlation in $^{14}$\textrm{O} $-$ $^{14}$\textrm{C} pair still exists and relatively stronger correlations without pairing effects are also observed. Moreover, relatively heavier mirror nuclei pairs show relatively good correlations between $\Delta R_{ch}$ and L. This analysis thus proves the robustness of these correlations and establishes the fact that pairing effects do weaken the correlation between the difference in charge radii of mirror nuclei and the slope of nuclear symmetry energy. Note that we do not see any clear systematics in mass effects on the Pearson Correlation coefficient as differences in mirror charge radii are also influenced by the neutron-proton asymmetry of the pair \cite{brow2}. 

\begin{figure}[!t] \centering \vskip 0.0cm
\includegraphics[scale=0.8]{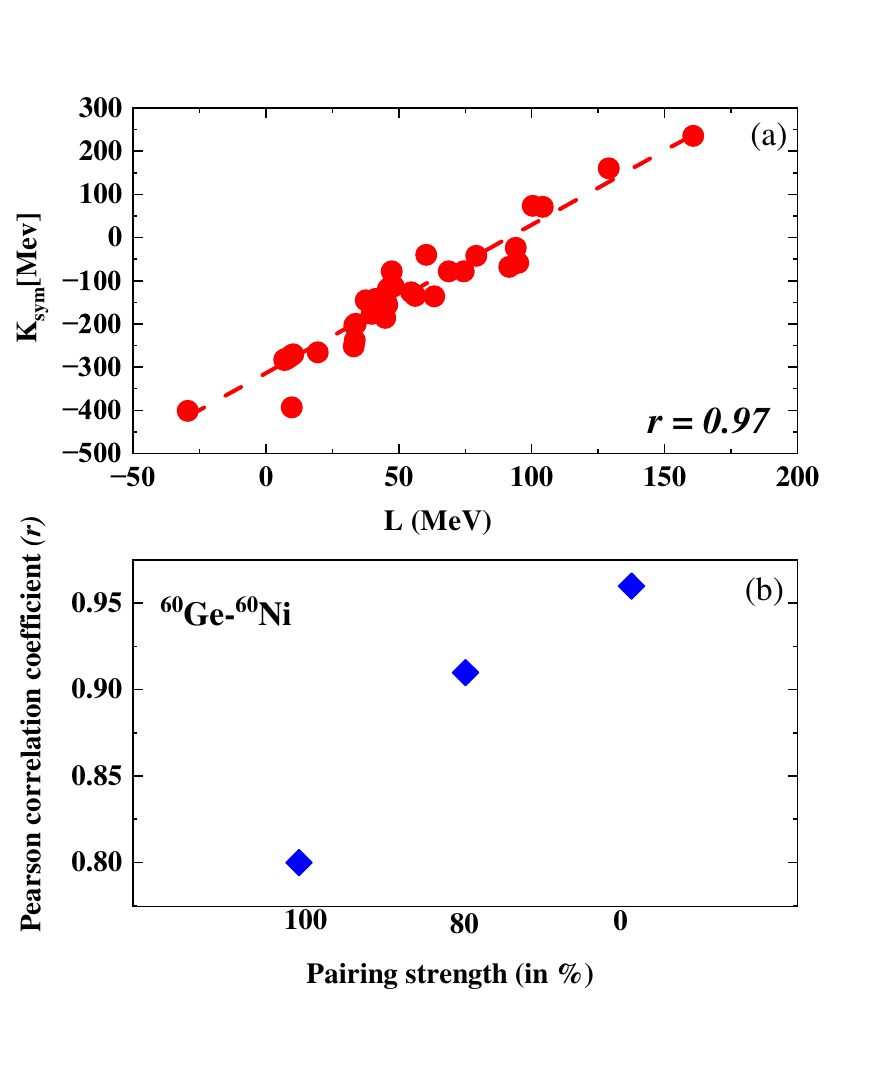}
\caption{(a) The Curvature of symmetry energy $K_{sym}$ as a function of the slope of symmetry energy L. Regression line is shown and a strong correlation is seen. (b) Pearson correlation coefficient for $\Delta R_{ch}$ and L correlation in the mirror nuclei pair of $^{60}\textrm{Ge}-^{60}$\textrm{Ni} for calculations with different pairing strengths.} \label{Fig3}
\end{figure}

\par
To understand the difference between the correlations obtained in the present study and Ref. \cite{huang}, we have shown the correlation between the curvature of symmetry energy and slope of symmetry energy in Fig. 3 (upper panel). We see that a relatively stronger correlation (of 0.97) is seen in the present study, which can change the correlations between $\Delta R_{ch}$ and L. Therefore, types of density functionals (Skyrme or relativistic) change the correlation between $\Delta R_{ch}$ and L as the curvature of symmetry energy influences this correlation, and thus different values of Pearson correlation coefficient are reported in the literature even for same mirror nuclei pair.
\par

Next, we also examine whether the inclusion of pairing effects systematically decreases the correlation or not. We performed the calculations for the heaviest pair of $^{60}$\textrm{Ge} $-$ $^{60}$\textrm{Ni}, where the best correlation is seen, with pairing effects implemented to their $80\%$ of the full strength. This is achieved by scaling the pairing strengths of neutrons and protons in different Skyrme forces accordingly and results are shown in Fig. \ref{Fig3} (bottom panel). From the figure, we see that the Pearson correlation coefficient obtained with $80\%$ pairing strength lies in between those obtained without pairing effects and with full pairing strength. Note that when pairing strengths are reduced below $80\%$, then pairing effects become too feeble for most of the Skyrme forces, to have any observable effect and results are close to those obtained without pairing force taken into account.

\begin{figure}[!t] \centering \vskip 0.0cm
\includegraphics[scale=0.75]{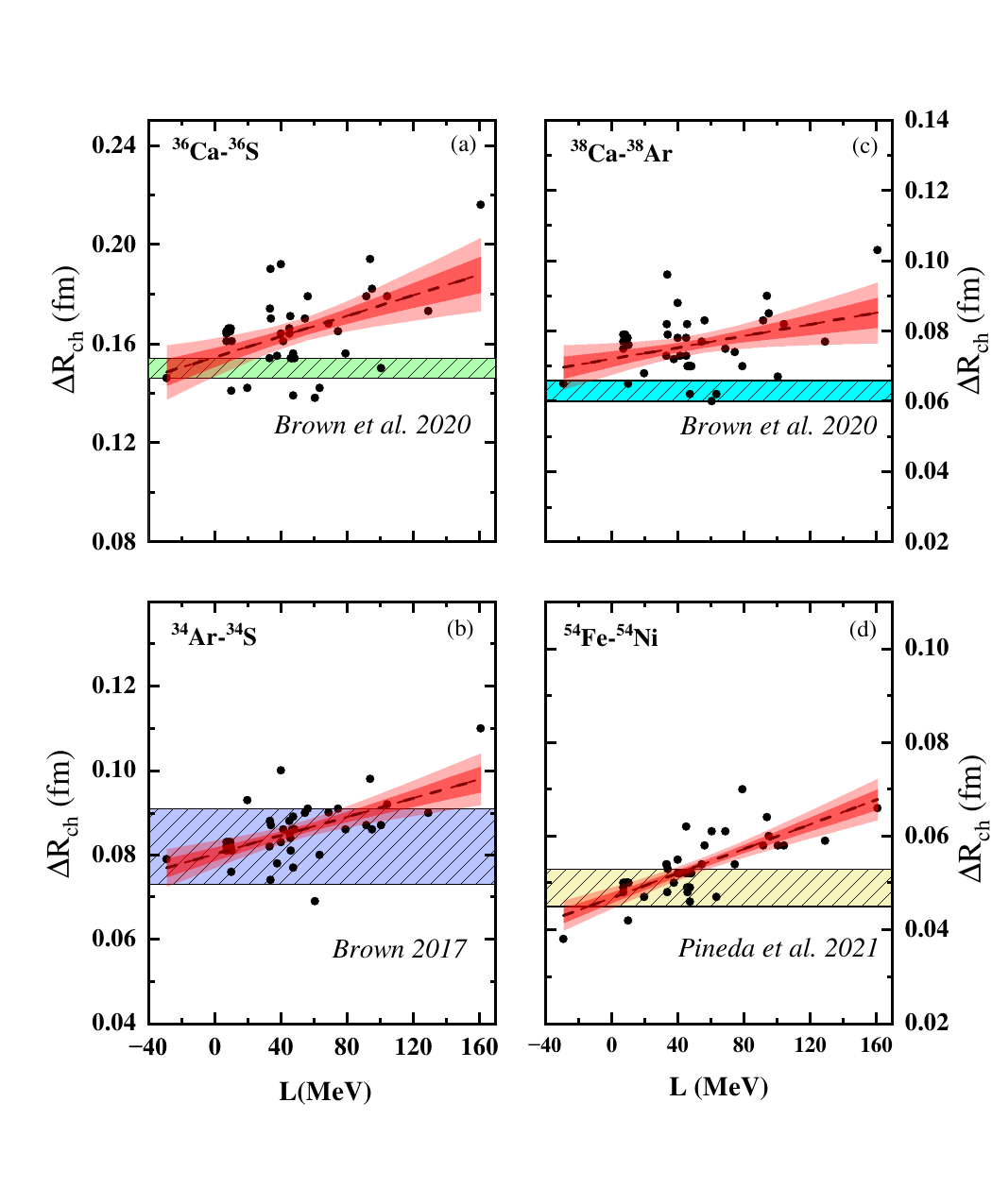}
\caption{$\Delta R_{ch}$  as a function of L for pairs of  $^{36}$\textrm{Ca} $-$ $^{36}$\textrm{S} [a], $^{34}$\textrm{Ar} $-$ $^{34}$\textrm{S} [b], $^{38}$\textrm{Ca} $-$ $^{38}$\textrm{Ar} [c],  and $^{54}$\textrm{Ni} $-$ $^{54}$\textrm{Fe} [d]. Experimental measurements are shown by horizontal bands and data is taken from Brown \emph{et al}. 2020 [a,c] \cite{brow3}, Brown 2017 \cite{brow2} [b], and Pineda \emph {et al.} 2021 [d] \cite{pine}. Dashed lines represent the least square linear fits and dark (light) shaded regions represent 68\% (95\%) confidence bands.} \label{Fig4}
\end{figure}
\begin{table}[ht]
\label{table:2}
\centering 
\caption {Experimental values of charge radii R$_{ch}$ and difference in charge radii $\Delta R_{ch}$ for mirror nuclei pairs of  $^{34}$\textrm{Ar} $-$ $^{34}$\textrm{S}, $^{36}$\textrm{Ca} $-$ $^{36}$\textrm{S}, $^{38}$\textrm{Ca} $-$ $^{38}$\textrm{Ar}  and $^{54}$\textrm{Ni} $-$ $^{54}$\textrm{Fe}. Values in the parentheses indicate systematic uncertainties.}
\begin {tabular} {|c|c|c|c|}
\hline 
\textrm {Nucleus} & \textrm {$R_{ch}$} & \textrm {$\Delta R_{ch}$} & \textrm {Ref.}\\
\hline
Ar$^{34}$ & 3.3657(21) &  & \\ 

S$^{34}$ & 3.284(2) & 0.082(9) & \cite{brow2}  \\
\hline
Ca$^{36}$ & 3.4484(27) &  & \\

S$^{36}$ & 3.2982(12) & 0.150(4)  & \cite{brow3} \\
\hline
Ca$^{38}$ & 3.4652(17) & & \\

Ar$^{38}$ & 3.4022(15) & 0.063(3)  & \cite{brow3}  \\
\hline
Ni$^{54}$ & 3.7370(30) & & \\

Fe$^{54}$ & 3.6880(17) & 0.049(4)  & \cite{pine}\\
\hline
\end{tabular}
\end{table}
\par
 Lastly, we attempt to obtain a constraint on the slope of nuclear symmetry energy from the experimental findings available on charge radius. Calculated values of $\Delta R_{ch}$ \emph{vs} L are plotted as solid circles in Fig. \ref{Fig4}. The measured values of charge radii are available only for a few nuclei and these pairs are  $^{34}$\textrm{Ar} $-$ $^{34}$\textrm{S} \cite{brow2}, $^{36}$\textrm{Ca} $-$ $^{36}$\textrm{S} \cite{brow3}, $^{38}$\textrm{Ca} $-$ $^{38}$\textrm{Ar} \cite{brow3} and $^{54}$\textrm{Ni} $-$ $^{54}$\textrm{Fe} \cite{pine}. The values of measured charge radii for different nuclei and the difference in charge radii are shown in Table \ref{table:2}. Experimental $\Delta R_{ch}$ values are displayed by horizontal bands in the figure. We observe that there is a very weak correlation seen in  $^{38}$\textrm{Ca} $-$ $^{38}$\textrm{Ar}, $^{36}$\textrm{Ca} $-$ $^{36}$\textrm{S} and $^{34}$\textrm{Ar} $-$ $^{34}$\textrm{S} pairs, with Pearson correlation coefficients lie between 0.3 and 0.5. Therefore, these pairs do not impose any noteworthy constraint on the slope of the nuclear symmetry energy. On the other hand, a slightly better correlation is seen in and $^{54}$\textrm{Ni} $-$ $^{54}$\textrm{Fe} pair, with a Pearson correlation coefficient of 0.75. Note that almost the same correlation is seen in Ref. \cite{rein2}, where the Coefficient of Determination is found to be 0.69 (which corresponds to a correlation coefficient close to 0.8) for $^{54}$\textrm{Ni} $-$ $^{54}$\textrm{Fe} pair, with HFB pairing case. It is worth mentioning that correlation is also significantly affected by the type of pairing scheme \cite{rein2}. Dark (light) shaded regions display 68\% (95\%) confidence bands. Therefore, $^{54}$\textrm{Ni} $-$ $^{54}$\textrm{Fe} pair imposes a relatively better constraint on the slope of the nuclear symmetry energy. The predicted values of L are $-20 \leq L \leq 55$ for 68\% confidence interval for $^{54}$\textrm{Ni} $-$ $^{54}$\textrm{Fe} pair. The maximum value of L becomes 62 MeV in 95\% confidence interval. Incidentally, this constraint is close to the L value reported recently by Lattimer \emph {et al.} \cite{latt} from combined analysis of neutron skin measurements and neutron star properties. Note that this value of L hints towards moderately soft symmetry energy and lies in between those estimated by CREX and PREX measurements, where soft and stiff forms of symmetry energy are inferred, respectively. The intermediate values of L are also proposed by Hu \emph {et al.} \cite{hu}, where calculations using ab-inito approach for nuclear forces from chiral EFT are performed. Their predicted slope parameter $L = 37-66$ MeV is found to be consistent with nuclear-scattering data as well. The above constraint is also in agreement with the one reported by S$\pi$RIT collaboration \cite{este}, where spectral pion ratio at high transverse momenta in Sn+Sn collisions is utilized to deduce the slope of the nuclear symmetry energy.  Further, this moderately softer form of nuclear symmetry energy is also in consensus with a recent study \cite{mour}, where EoS is constrained using observational data on neutron stars and low-mass X-ray binaries. An optimal range of L is found to be $50.79^{+15.16}_{-9.24}$ MeV in 68\% confidence interval when Bayesian analysis is done using priors chosen from PREX-II measurements \cite{reed,este,essi,yue}. Therefore, from our analysis, we conclude that the more accurate difference in the charge radii in heavier mirror nuclei pairs provides an alternative and independent probe to constrain the slope of the symmetry energy. Thus, new measurements on the charge radii of relatively heavier nuclei, along with neutron-skin measurements will be crucial for obtaining a much more stringent constraint on the slope of nuclear symmetry energy.  

\section{SUMMARY}
Correlations between the difference in charge radii of mirror nuclei ($\Delta R_{ch}$) and the slope of nuclear symmetry energy (L) are revisited by invoking pairing effects in the calculations. The said correlation is calculated for mirror nuclei pairs of $^{14}$\textrm{O} $-$$^{14}$\textrm{C}, $^{18}$\textrm{Ne} $-$$^{18}$\textrm{O}, $^{44}$\textrm{Cr} $-$ $^{44}$\textrm{Ca}, $^{58}$\textrm{Zn} $-$ $^{58}$\textrm{Ni} and $^{60}$\textrm{Ge} $-$ $^{60}$\textrm{Ni} using 40 Skyrme energy density functionals. It is seen that $\Delta R_{ch}$ is better correlated with L in heavier nuclei relative to those in lighter counterparts, with much stronger correlations obtained when pairing effects are not taken into consideration. These features remain preserved when density functionals are constrained for nuclear binding energy which reflects the robustness of these correlations. An attempt to derive a feasible value for slope of symmetry energy is made by comparing the calculations on charge radii differences with measurements of $^{34}$\textrm{Ar} $-$ $^{34}$\textrm{S}, $^{36}$\textrm{Ca} $-$ $^{36}$\textrm{S}, $^{38}$\textrm{Ca} $-$ $^{38}$\textrm{Ar},  and $^{54}$\textrm{Ni} $-$ $^{54}$\textrm{Fe} pairs. Our constraint from measured $\Delta R_{ch}$ hints at moderately soft symmetry energy, with 
predicted values of L are -20 MeV $\leq L\leq$ 55  MeV with 68\% confidence band, which is in harmony with recent studies done on simultaneous compatibility of neutron skins reported in CREX and PREX-II data and astrophysical observations on neutron stars.
Further, our study predicts that new measurements on the charge radii of relatively heavier nuclei will be fruitful in achieving independent constraints on the slope of the nuclear symmetry energy. Such constraints can then be complemented with measurements of neutron skin to impose much better information about the behavior of nuclear symmetry energy. 
 \section{Acknowledgements}
AV would like to acknowledge the CSIR for the support through the CSIR-JRF 09/1026(16303)/2023-EMR-I. SG partially acknowledges CSIR for the support vide. grant no. 03/1513/23/EMR-II.

\end{document}